\documentclass[aps,prl,twocolumn,groupedaddress]{revtex4}

\usepackage{graphicx}
\usepackage{dcolumn}
\usepackage{bm}

\begin{document}


\title{Quantum Monte Carlo simulations for the Bose-Hubbard model with
random chemical potential; localized Bose-Einstein condensation without superfluidity}


\author{
Mitsuaki Tsukamoto and Makoto Tsubota
}
\affiliation{
Department of Physics, Osaka City University, Sumiyoshi-Ku, Osaka
558-8585, Japan
}


\date{\today}

\begin{abstract}
The hardcore-Bose-Hubbard model with random chemical potential is
investigated using quantum Monte Carlo simulation.
We consider two cases of random distribution of the chemical potential:
a uniformly random distribution and a correlated distribution.
The temperature dependences of the superfluid density,
the specific heat, and the correlation functions are calculated.
If the distribution of the randomness is correlated, there exists an intermediate state, which can be thought of as
a localized condensate state of bosons, between the superfluid state
and the normal state.
\end{abstract}


\maketitle

Bose-Einstein condensation and superfluidity are two important concepts in modern physics. 
Both of these concepts are characteristic of the ground states of quantum fluids
and have been investigated in lots of types of systems. 
However, the relation between Bose-Einstein condensation and superfluidity remains unclear\cite{BECbook}. 
Bose-Einstein condensation was first predicted by Einstein for an ideal Bose
gas. The Bose-Einstein condensate (BEC) was later observed as the $\lambda$ transition
of strongly-correlated liquid $^4$He, 
and has recently been realized in dilute atomic gases\cite{PethickSmith} 
and excitons in semiconductors\cite{Exciton}.  
For non-interacting particles, the criterion for Bose-Einstein condensation 
is that the occupation number for one of the single-particles energy levels should be macroscopic. 
The general criterion for interacting systems is that the one-particle density
matrix $\rho (\mathbf{x}, \mathbf{y})=\langle \widehat{\psi}^{\dagger}(\mathbf{x})\widehat{\psi}(\mathbf{y}) \rangle$ 
tends to a constant as  
$|\mathbf{x}-\mathbf{y}| \rightarrow \infty$. 
In contrast, the term superfluidity originally described a group of experimental phenomena such as frictionless flow and persistent current. 
After the observation of superfluidity in liquid $^4$He, superfluidity was observed in Fermionic liquid $^3$He and electrons in metals as superconductivity, and is probably realized even in neutron stars.  
As proposed by Hohenberg and Martin\cite{HohenbergMartin}, 
superfluidity is not an equilibrium property of the system, but rather an anomaly of the transport coefficient. However, our understanding of superfluidity is considerably less than our understanding of Bose-Einstein condensation. 
Our first system showing Bose-Einstein condensation and superfluidity is liquid $^4$He, in which Bose-Einstein condensation and superfluidity occur simultaneously at the $\lambda$ temperature, but these two concepts are, in principle, not equivalent. 
Recent studies on disordered Bose systems, such as $^4$He in restricted geometry, describe the separation of BEC and superfluidity. Furthermore, the nature of both Bose-Einstein condensation and superfluidity has been investigated. 
 The present study is an attempt to solve the essential problem in physics  by quantum Monte Carlo simulation.

The disordered Bose system is actualized for liquid 
$^4$He confined in the
nano-porous glass\cite{Reppy1,Reppy2,Vycor,Shirahama} and the atomic gas in the optical trap potentials\cite{Opt}.
In particular, the confinement of $^4$He has been investigated experimentally
using various porous glasses, such as 
Vycor\cite{Vycor},  Aerogel\cite{Aerogel}, and Gelsil\cite{Shirahama}.
In addition, the confinement of the $^4$He 
in a uniform-nano-porous medium has also attracted interest\cite{Wada,Wada2}.
The pressure-temperature phase diagrams change drastically depending on the dimensionalities and pore-sizes of the medium.
The superfluidity is suppressed as the pore-size becomes narrow, 
because the interaction between $^4$He atoms increases with 
confinement in small pores and applied  pressure.
For the $^4$He in the FSM\cite{Wada}, which is one-dimensional and uniform porous, there exists a critical pressure above which
the normal-fluid state is maintained and  the superfluidity does not appear, even at zero
temperature, if the diameter of the pore size is less than 2.8 nm.

Recently, low-temperature experiments \cite{Shirahama} using 
$^4$He in nano-porous Gelsil glass, which has a mean pore size of 2.5 nm, 
revealed that the onset temperature of the superfluidity was lower than  the peak temperature of the specific heat that corresponds to the transition temperature to the 
BEC state.
This means that between the SF state and the normal fluid state, an intermediate state, called the localized Bose-Einstein condensate (LBEC) state, exists.
In the LBEC state, the $^4$He is in the BEC state,
but superfluidity cannot be detected through a torsional oscillator experiment. 
The LBEC state can be thought of as a localized state of the BECs
by the random potential created by the nano-porous medium. 
Note that for the case of
the $^4$He in the uniform medium HMM3\cite{Shirahama2}, which has the same pore size as the mean size of the Gelsil glass,
there is no clear indication of the existence of the LBEC state. 
In addition, the dependences of the superfluidity and the LBEC state on 
the randomness and pore-size of the porous medium are of particular interest.

The above described experimental results suggest the problem  
of how the superfluidity depends on the randomness and the finite-size effect of the confinement. In addition, $^4$He in the three-dimensional system of Gelsil glass has different transition temperatures between the BEC and the SF state. This can be realized by cooperation among the randomness and the finite-size effect. This system differs completely from the three-dimensional bulk $^4$He case, in which the Bose-Einstein condensation and SF transition occur simultaneously at the $\lambda$ temperature. 


In the present paper, we investigate the randomness 
and its distribution dependence on the Bose system.
We consider the Bose-Hubbard model with random chemical potential.
Taking a hard-core limit for simplicity, the model Hamiltonian 
on the three-dimensional cubic lattice with periodic boundary condition is given as
\begin{eqnarray}
 {\cal H} = -J \sum_{<i,j>} (\hat b^{\dagger}_i \hat b_j + {\rm H.C.})
  + \sum _i \mu _i \hat n_i ,
\label{Hamil}
\end{eqnarray}
where $\hat b_i^{\dagger}$($\hat b_i$) is a boson creation (annihilation)
operator, $\hat n_i = \hat b_i^{\dagger} \hat b_i $, and
$\mu_i$ is the random potential that depends on site $i$.
Here, we define $\mu_i$ such that
it can takes two values, $-\mu_{-}$ and $+\mu_{+}$ ($\mu_{-},\mu_{+} >0 $).
For the random distribution of $\{ \mu_i \}$,
we consider the two cases. 
One case is a uniformly random distribution, that is, the sites at which 
$\mu_i = \mu_{+}(-\mu_{-})$ are uniformly distributed with probability 
$P_{+}(P_{-}=1-P_{+})$.
The other case is a correlated distribution, that is, 
the sites at  $\mu_i = -\mu_{-}$ tend to make clusters, and
the clusters of $-\mu_{-}$ are placed randomly 
such that the density of the sites of $-\mu_{-}$ is $P_{-}$.
We set the mean size of the clusters to seven to eight sites.
We perform quantum Monte Carlo simulation using a directed loop
algorithm\cite{dla1} 
in order to investigate the Hamiltonian (\ref{Hamil}) and calculate the temperature dependences of the susceptibility, the  SF density, the specific heat, and the two-point correlation function.
The random average for the chemical potential is taken over more than 1,024 samples.
The dilute Bose system\cite{Kobayashi} and the mean-field approach\cite{Krutitsky} have been investigated in terms of the uniformly distributed random potential at finite temperature. The thermodynamic property of the Hamiltonian (\ref{Hamil}) has not been investigated directly.

\begin{figure}[ht]
 \includegraphics[clip,scale=0.32,angle=270]{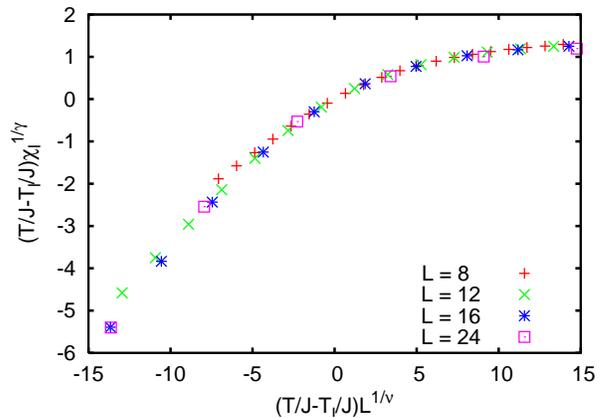}
 \caption{(color online). Finite-size-scaling plot of the $\chi_{\rm l}$
 using the critical exponents of the three-dimensional XY model,
 $\gamma=1.3177$ and $\nu = 0.67155$.
 The data of the systems of different sizes follow the same curves,
 which means that FSS works well. We estimate the transition temperature 
 $T_{\rm l}$ to the locally BEC state in this manner.
 The error bars of the data are smaller than the size of the symbols. }
 \label{FSS}
\end{figure}

Here, we introduce two susceptibilities to
determine the  transition temperatures corresponding to the two states:
the state in which the bosons at $\mu_i = -\mu_{-}$
form BEC clusters and the SF state over the entire system.
Note that, according to Harris's criterion\cite{Harris},
the universality class of the three-dimensional XY model is not affected by on-site potential. Thus, we can determine the critical temperature 
from the finite-size scaling (FSS) using the critical exponents of the three-dimensional XY model. We compute the system size dependence of the 'total' susceptibility as follows:
\begin{eqnarray}
\chi _{\rm t}(L,T) = \frac{1}{\beta N} \sum_{i=1} ^N 
\sum_{r}
\int ^{\beta}
\langle b_i^{\dagger}(0)b_{i+r}(\tau)\rangle   d\tau ,
\label{chx}
\end{eqnarray}
where $\tau$ is the  imaginary time. 
We then perform FSS.
The critical temperature obtained from the FSS of the $\chi_{\rm t}$ 
corresponds to the onset temperature of the SF
density because the correlation length extends over the entire system 
at the temperature. 
We call this temperature the critical temperature $T_{\rm t}$,
at which the entire system is in the SF state.
Next, we  define the local BEC order' and its susceptibility, as follows: 
\begin{eqnarray}
 \chi_{\rm l}(L,T) = \frac{1}{\beta N_{-}}\sum_{i\in \{-\}}
  \sum_r
  \int ^{\beta}\langle b_i^{\dagger}(0)b_{i+r}(\tau)\rangle d\tau ,
\label{chx_l}
\end{eqnarray} 
where $N_{-}$ is the number of sites at which $\mu = -\mu_{-}$
and $\sum_{i\in \{-\}}$ is a limited summation over these sites.
Equation (\ref{chx_l}) is an integration of the correlation 
between a site at which $\mu = -\mu_{-}$ and a site that is located at a distance $r$ distant the $\mu = -\mu_{-}$ site.
From the FSS of the $\chi_{\rm l}$, we define the critical temperature 
$T_{\rm l}$ at which  clusters of the $\mu_i = -\mu_{-}$ 
sites are locally in the BEC state.

Figure \ref{FSS} shows the FSS plot of $\chi_{\rm l}$ 
when $\mu_{-} = 2.0$, $\mu_{+} = 7.0 $, and $P_{-} = 0.2$, and the random distribution has a correlation.
Finite-size scaling using the exponents of the three-dimensional XY model is found to work
well for the case in which $T_{\rm l} / J = 0.53$.
In contrast, the FSS of the  $\chi_{\rm t}$ yields the critical temperature 
$T_{\rm t} /J = 0.49$. For the case in which the random potential is distributed uniformly, there is no difference between $T_{\rm l}$ and $T_{\rm t}$
within the numerical error. For the case of the correlated random distribution, 
the bosons at $\mu_i = -\mu_{-}$ locally enter the BEC state
at $T_{\rm l}$ and form clusters, which then link together as the temperature decreases. When the locally correlated clusters percolate 
through the entire system, superfluidity occurs at $T_{\rm t}$.
Since we are interested in  the intermediate state between $T_{\rm l}$ and $T_{\rm t}$, in the remainder of the present paper, we discuss only the case in which the 
distribution of the random potential has a correlation.

\begin{figure}[ht]
(a) 
 \includegraphics[clip,scale=0.3,angle=270]{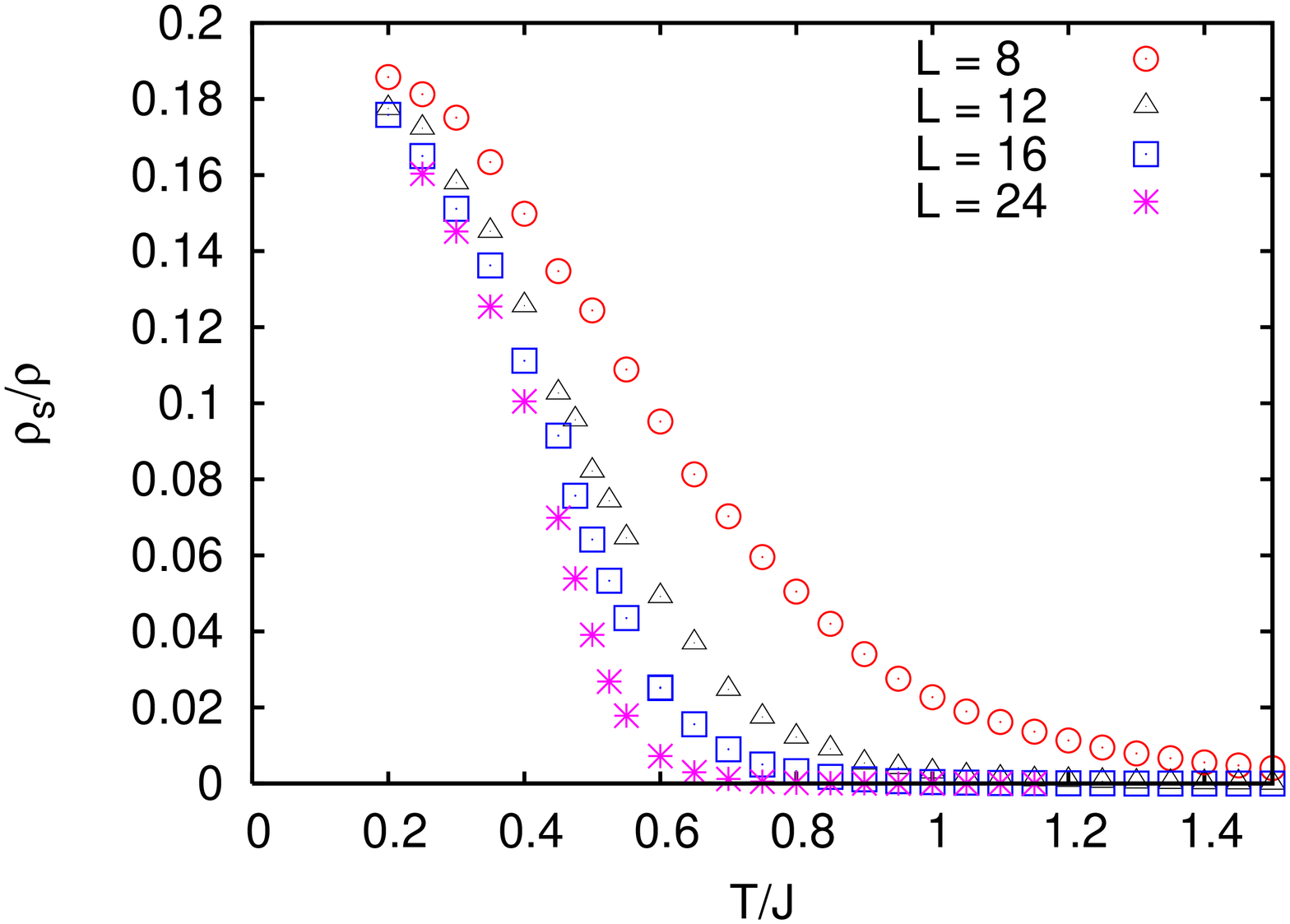}

(b) 
 \includegraphics[clip,scale=0.3,angle=270]{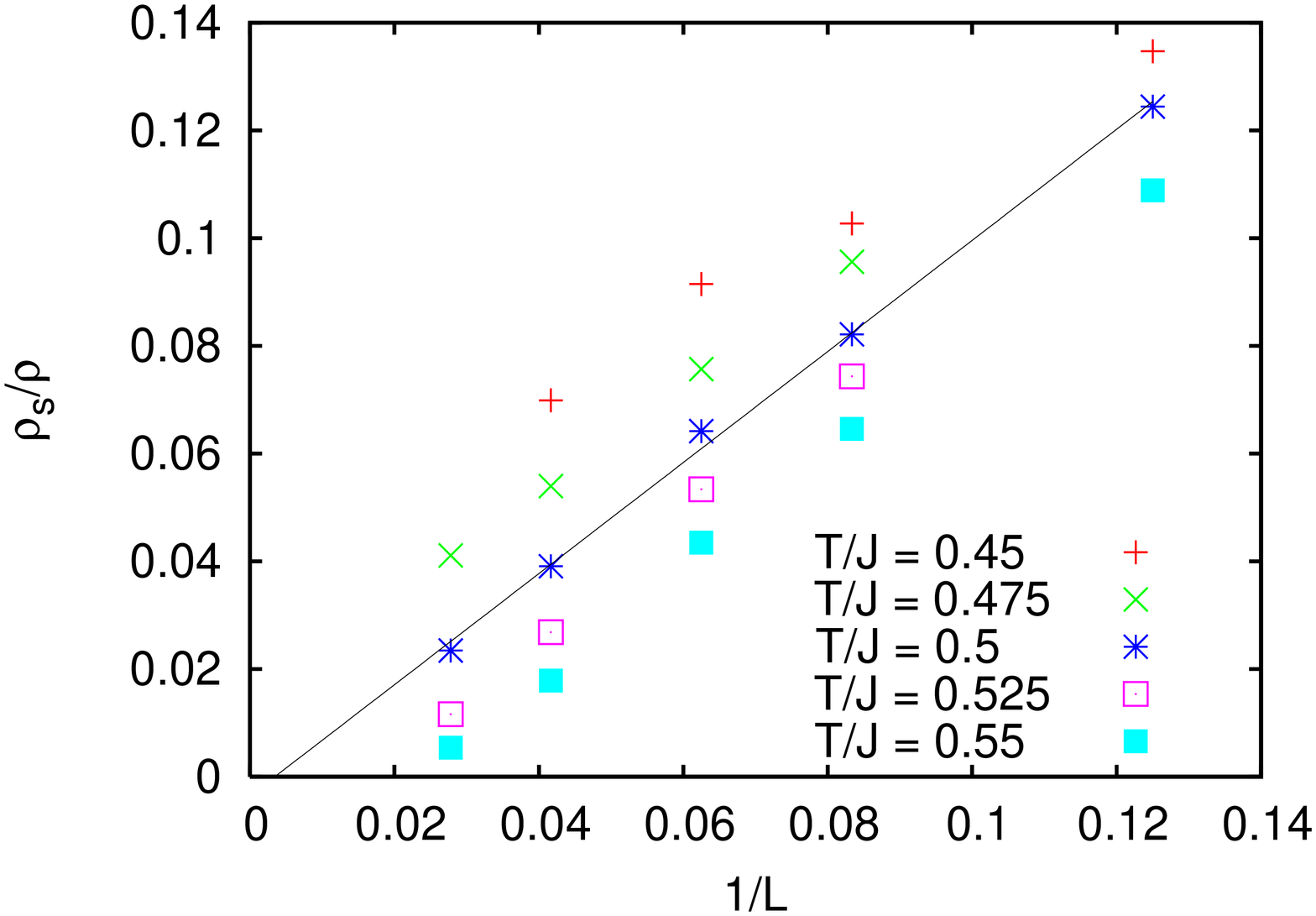}
 \caption{(a) Temperature dependence and (b) the system size dependence of the
 SF density for the case in which $\mu_{-} = 2.0$, $\mu_{+} = 7.0 $, and $P_{-} = 0.2$. The temperatures estimated from FSS for $\chi_{\rm t}$ and 
 $\chi_{\rm l}$ are $T_{\rm t}/J = 0.49$ and $T_{\rm l}/J = 0.53$, respectively. 
 The onset temperature of the superfluidity is consistent with $T_{\rm t}$. The
 line in (b) is a fitting plot of $T/J = 0.5$. (color online).
 }
\label{hel}
\end{figure}
Figure \ref{hel}(a) shows 
the temperature dependence of the SF density \cite{SuperFluid}
for various system sizes, and
 Fig. \ref{hel}(b) shows the SF density as a function of the inverse
system size $1/L$. The line in the Fig. \ref{hel}(b) is a fitting plot of
$T/J=0.5$. The onset temperature of the superfluidity corresponds to the
critical temperature $T_{\rm t}$. 
Hence we find that below $T_{\rm t}$ both BEC and superfluidity are achieved
over the entire system.
We consider the intermediate state of the temperature region between  
$T_{\rm l}$ and $T_{\rm t}$ as the state in which the BECs are localized by
correlated random chemical potential 
the entire system does not show the superfluidity.
 
\begin{figure}[ht]
 \includegraphics[clip,scale=0.3,angle=270]{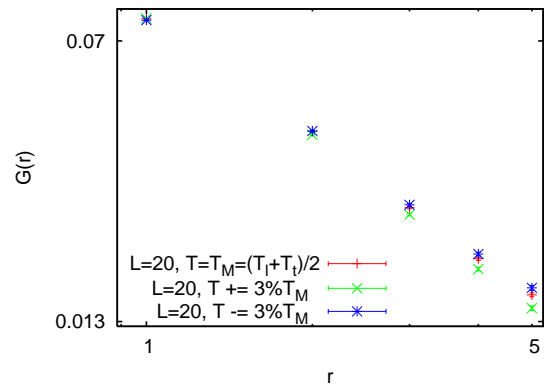}
 \caption{Two-point correlation function as a function of 
 distance(logarithmic plot). The error bars of the data are smaller than the symbols in the figure. (color online).}
 \label{Corr}
\end{figure}
In Fig. \ref{Corr}, we plot the two-point correlation function
\[
 G(r) = \frac{1}{N}\sum_i \langle b_i^{\dagger}b_{i+r}\rangle,
\]
as a function of the distance $r$.
We calculate $G(r)$ for system sizes of $L=16$ and $20$.
Comparing the results for $L=16$ and $20$, we find that
if $r<6$, there is no difference between the results. Therefore, in order to negate the finite-size effect, we plot only the region of $r<6$.
For the ordinary second-order transition belonging to the three-dimensional XY
universality class, we have $G(r) \propto 1/r^{1+\eta} = 1/r^{1.038}\, $
\cite{3DXY} at the critical temperature, above (below) which the $G(r)$ decreases exponentially (converges to some constant value) as 
$r \rightarrow \infty$. In the present case, however, the power law behavior is observed for a finite temperature range. Because of the random potential, the critical behavior is unclear, and there appears to be a finite critical region.
Since this critical region is small, however, FSS analysis is feasible.

\begin{figure}[ht]
 \includegraphics[clip,scale=0.3,angle=270]{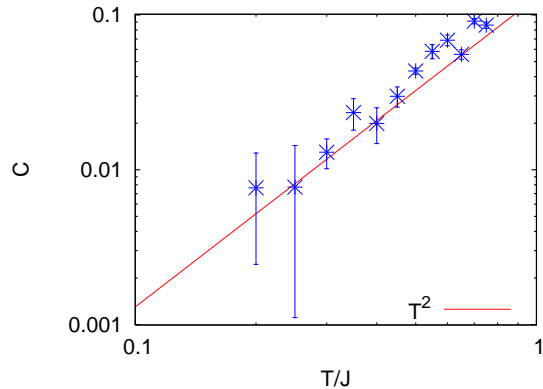}
 \caption{ Logarithmic plot of the temperature dependence of the specific heat.
 The line in the figure is a fitting plot of $C \propto T^2$.(color online).
}
 \label{Spc}
\end{figure}
The temperature dependence of the specific heat is shown in Fig. \ref{Spc}.
The line is a curve-fitting plot in which 
\[
 C \sim T^2.
\]
The data is well fitted by the line, while for the uniform system $\mu_i = \mu$, we find the specific heat behaves as $C \sim T^3$ at low temperature as a result of the gapless and linear dispersion of the elementary excitations. The $T$-square behavior of the specific heat is also observed in the experiment to examine the $^4$He in the Gelsil glass\cite{Shirahama2}. However, it is thought that $T$-square behavior in the experiment is a cross-over of the $T$-linear behavior, which is characteristic of the glass state. Thus, it is expected that $C\sim T$ as $T\to 0$\cite{Shirahama2}.

\begin{figure}[ht]
(a)
 \includegraphics[clip,scale=0.32,angle=270]{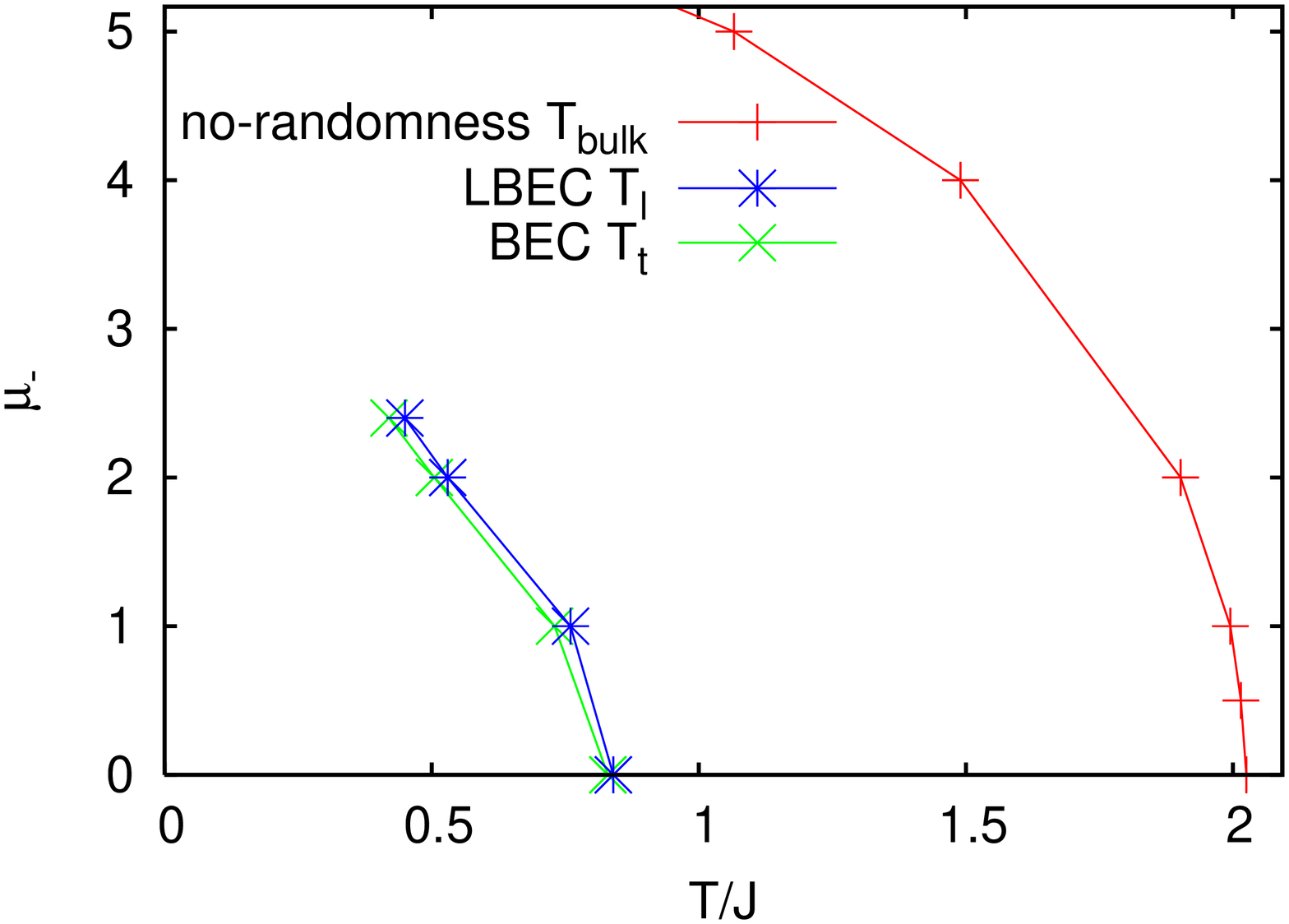}

(b) 
\includegraphics[clip,scale=0.3,angle=270]{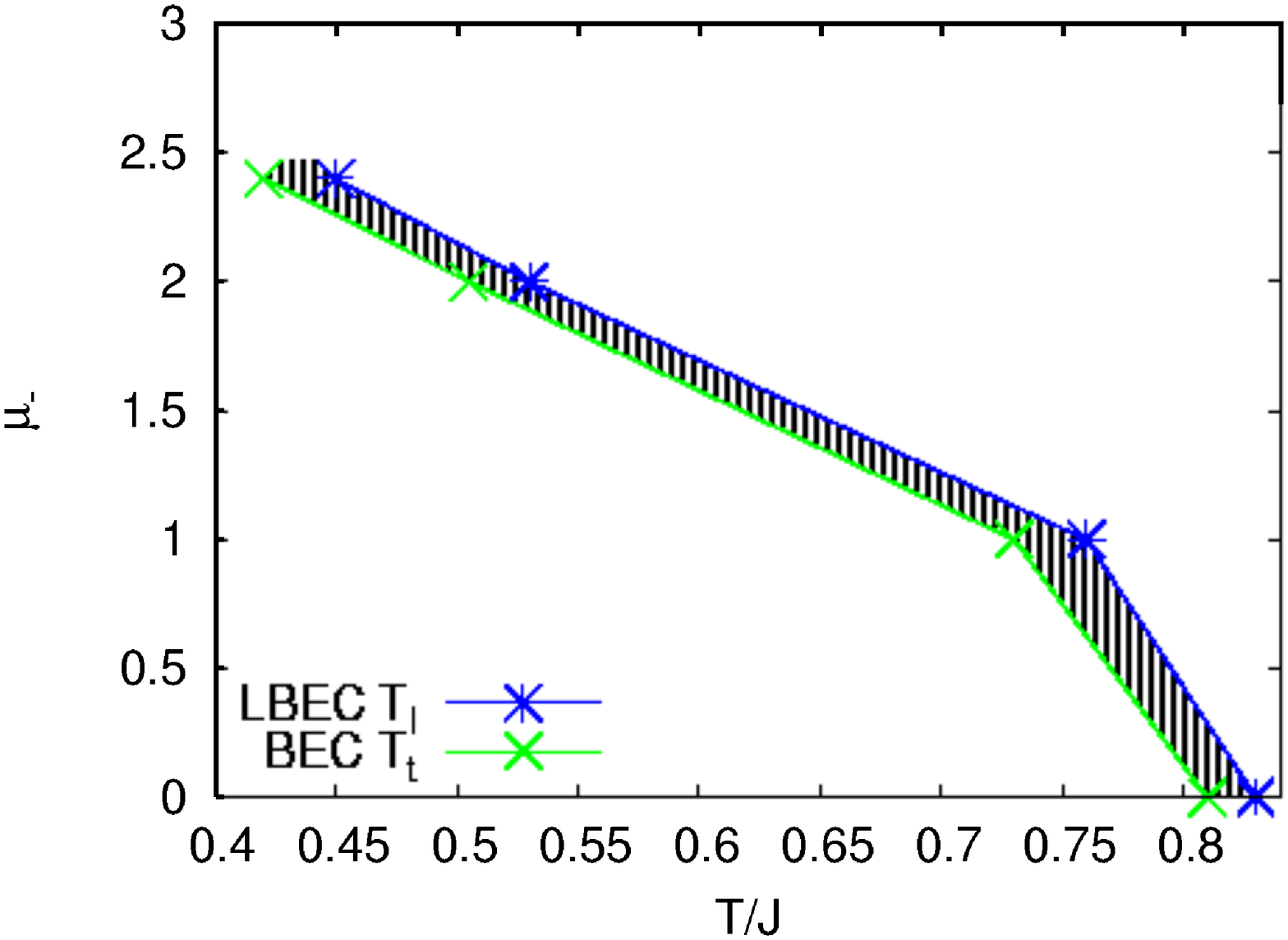} 
 \caption{
Phase diagram in the $\mu _{-}$ - $T$ plane.
 (a) An overview of the phase boundaries $T_{\rm bulk}$,
 $T_{\rm l}$, and $T_{\rm t}$.
 The phase boundaries of the system with
 randomness and non-randomness are approximately parallel.
(b) An enlarged view of the LBEC region.
 The vertical-striped region is the state where the bosons are locally condensed.
 The error bars of the data are smaller than the size of the symbols.
 (color online).
}
 \label{Phase}
\end{figure}

Figure \ref{Phase} shows the phase diagram of the $\mu_{-} - T$ plane
for $\mu_{+} =7.0 $ and $P_{-}=0.2$. 
In the phase diagram, in addition to  $T_{\rm t}$ and $T_{\rm l}$, we plot the critical temperature $T_{\rm bulk}$ of the system with no-randomness.
Comparing the system with randomness to that without randomness,
$T_{\rm t}$ is found to be lower than $T_{\rm bulk}$, which means that
the superfluidity is suppressed because of the chemical potential $\mu_{+}$.
Comparing $T_{\rm t}$ and $T_{\rm l}$, $T_{\rm t}$ is always lower than
$T_{\rm l}$ at the same $\mu_{-}$.
In the vertical-striped region of Fig. \ref{Phase}(b), bosons are locally condensed but
the entire system is not in the SF state.
Note that the phase boundaries of the system with no-randomness, $T_{\rm t}$ and $T_{\rm l}$, are approximately parallel.

In conclusion, we have investigated a Bose-Hubbard model with random chemical
potential by quantum Monte Carlo simulation. If the distribution of the random potential has correlation, the localization of the BECs occurs at a temperature  
above the onset temperature of the superfluidity. We have obtained the phase diagram of the $\mu _{-}$-$T$ plane. In the intermediate state, with respect to the distance, the two-point correlation function displayed power-law behavior.

The reader may wonder whether the locally condensed state obtained in the present 
study does not correspond exactly to the experimentally observed LBEC state\cite{Shirahama}. In the present study, 
the bosons in the cluster of the $\mu_{i} = -\mu_{-}$ first locally enter the BEC state, and the superfluidity for the entire system then grows as a result of 
quantum fluctuations as the temperature decreases. 
On the other hand, the experiment involving the $^4$He in the various
nano-porous media, the critical temperature depends on the pore size because of
the confinement, where the $^4$He is strongly correlated under the
pressure. Each cluster has a different critical temperature depending on the
strength of the correlation. As a result, the SF state is experimentally
achieved by percolation through the fraction of such LBECs. In the present
study, an intermediate state exists between the locally condensed state and the
SF state. However, the present situation is not equivalent to that of the
experiments, because the correlation of the boson is too strong owing to the hard-core limit.

\bibliography{basename of .bib file}

\end{document}